\documentclass{svjour3}                     
\smartqed  
\usepackage{graphicx}
\usepackage{mathptmx}      
\journalname{Journal of Low Temperature Physics}
\begin{document}

\title{Statistics of Intermittent Switching between Potential Flow and Turbulence around an Oscillating Sphere in Superfluid $^4$He below 0.5 K}


\titlerunning{Statistics of Intermittent Switching}        

\author{W. Schoepe        
}


\institute{W. Schoepe \at
              Fakult\"at f\"ur Physik, Universit\"at Regensburg, Germany \\
              \email{wilfried.schoepe@physik.uni-regensburg.de}           
}

\date{Received: date / Accepted: date}

\maketitle

\begin{abstract}
The flow of superfluid helium at very low temperatures around an oscillating microsphere is known to be unstable slightly above the critical velocity. The flow pattern switches intermittently between potential flow and turbulence. From time series recorded at constant temperature and driving force the statistical properties of the switching phenomenon are discussed. Based on our recent understanding of the critical velocity $v_c$ of oscillatory superflows, i.e., $ v_c \sim \sqrt{\kappa \,\omega}$, where $\kappa$ is the circulation quantum and $\omega$ is the oscillation frequency, the analysis is being refined now. From the exponential distribution of the lifetimes of the turbulent phases quantitative information on the vortex density $L$ can be inferred such as the distribution and the width of the fluctuations of $L$. The phases of potential flow show a Rayleigh distribution of the excess oscillation amplitude above the amplitude at turbulent flow. The rms value is found to scale as $\omega^{-3/2}$.\\  

\keywords{Quantum turbulence \and Oscillatory flow \and Intermittent switching}
\PACS{67.25.Dk \and 67.25.Dg \and 47.27.Cn}
\end{abstract}

\section{Introduction}
\label{intro}
Laminar and turbulent flow of superfluid $^4$He around an oscillating microsphere (radius r = 0.12 mm) has been studied in detail in recent years \cite{PRL1,OVL}. At small oscillation amplitudes the drag force is linear in velocity amplitude (Stokes regime) and is determined by the normal fluid component. At temperatures below ca. 0.5 K the drag is given solely by ballistic phonon scattering and vanishes as $T^4$. Above a critical velocity amplitude $v_c$ a transition to a large and nonlinear drag force is observed that scales as $(v^2 - v_c^2)$, see Fig.1. (The small remaining phonon drag can subtracted from the total drag force and will not be considered here any more.) This transition is interpreted as the onset of turbulent flow of the superfluid. For velocities just above $v_c$ an instability of the flow is observed at temperatures below ca. 0.5 K, namely an intermittent switching of the flow pattern between turbulent and laminar phases \cite{OVL}.\\ 

\begin{figure}
\includegraphics{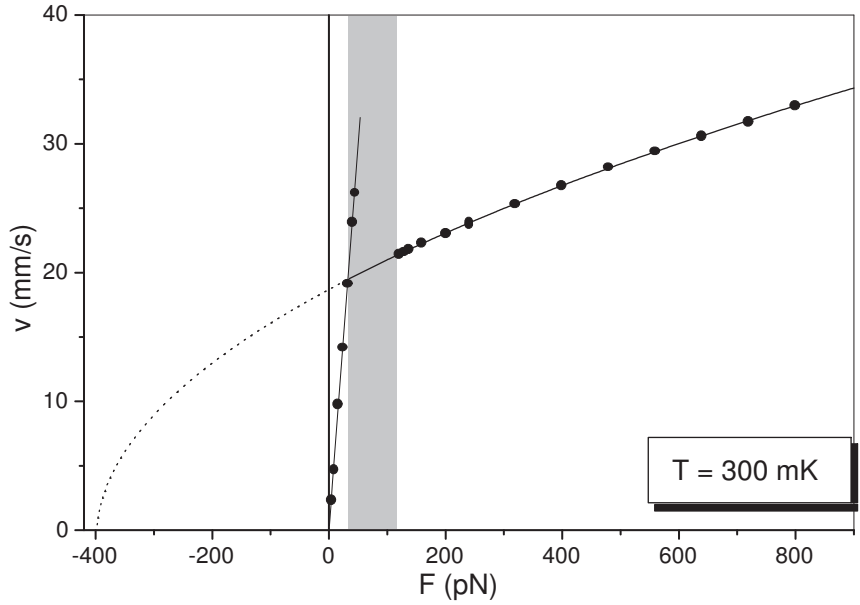}
\caption{Velocity amplitude $v$ of the oscillating sphere (frequency 114 Hz) as a function of the driving 
force $F$ at $300\,\mbox{mK}$. At small drives the flow is laminar and the linear 
behavior is determined by ballistic phonon scattering. The shaded area above the critical velocity $v_c \approx $ 20 mm/s indicates the unstable regime where the flow switches intermittently between potential flow and turbulence. At larger driving forces we observe stable nonlinear turbulent drag where $F(v)$ scales as $(v^2 - v_c^2)$.}

\label{fig:1}       
\end{figure}

In Fig.2 time series recorded at constant temperatures and at three different driving forces are shown. During a turbulent phase the drag is given by the turbulent drag force and, hence, the velocity amplitude is low. It is obvious that the lifetime of the turbulent phases grows very rapidly with the driving force. When turbulence breaks down the velocity amplitude begins to grow due to the much smaller phonon drag. Because the velocity amplitudes are above $v_c$ these phases of laminar flow (better: potential flow) usually break down and a new turbulent phase is observed. In some cases, in particular at low drive, metastable potential flow with very long lifetimes is observed occasionally when the maximum amplitude corresponding to phonon drag is reached (see upper trace in Fig. 2), as will be discussed later.\\

The statistical properties of this switching phenomenon have been analyzed in some detail already in earlier work \cite{OVL,PRL2}. Since then, our understanding of the transition to turbulence of oscillatory flow at a critical velocity $v_c$ has been greatly improved by identifying a universal critical velocity that does not depend on any geometry, namely $v_c\sim\sqrt{\kappa \cdot \omega}$, where $\kappa$ is the circulation quantum and $\omega$ is the oscillation frequency \cite{JLTP153,JLTP158}. This result can be derived from a general argument based on the ``superfluid Reynolds number'' and in more detail from the dynamical behaviour of the turbulent vortex tangle as recently discussed by Kopnin \cite{JLTP153,kolya}. It is the purpose of this work to present now a more refined analysis of the statistics of this switching phenomenon. In the following Section 2 the tools necessary for the statistical analysis will be outlined, in Section 3 the analysis of the statistics of the turbulent phases will be discussed. It will be shown that quantitative information on the turbulent vortex density $L$ and its fluctuations can be inferred from the data. Moreover, it will be emphasized again that superfluid turbulence is inherently unstable and only appears to be stable due to the finite time of experimental observation. In Section 4 the statistical properties of the phases of potential flow will be presented. Finally, the observation of metastable potential flow above $v_c$ will be discussed. Section 5 concludes this work.\\


%

\begin{figure*}
  \includegraphics[width=1\textwidth]{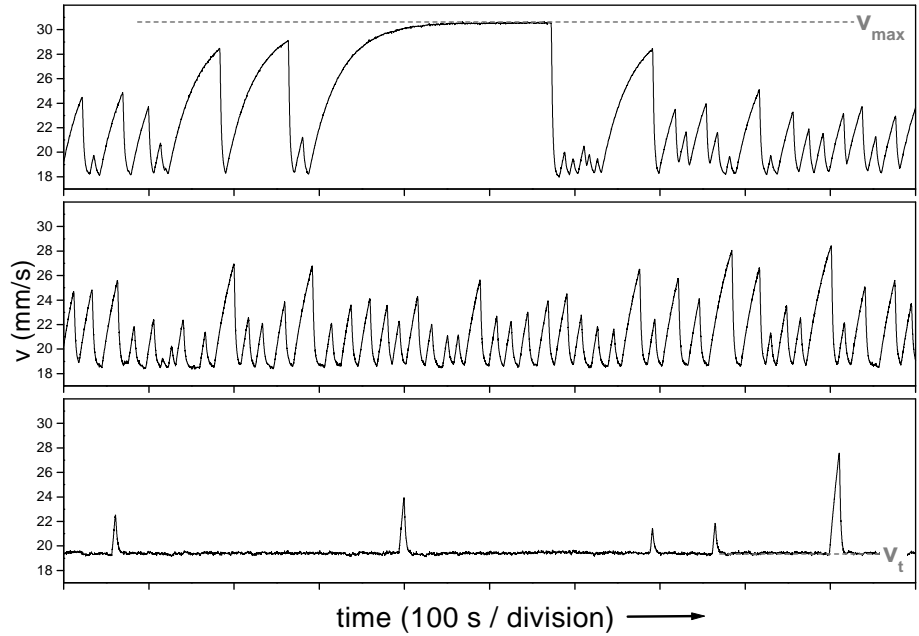}
\caption{Three time series of the velocity amplitude at 300 mK and at three different driving forces (in pN): 47, 55, 75 (from top to bottom). The low level $v_t$ corresponds to turbulent flow while the increase of the velocity amplitude occurs during a laminar phase, occasionally reaching the maximum value $v_{max}$ given by the linear drag. With increasing drive the lifetimes of the turbulent phases grow rapidly.}
\label{fig:2}       
\end{figure*}
%
\section{Statistical Tools}
\label{sec:2}

As presented earlier \cite{OVL} the reliability function $R(x)$ is the relevant quantity of our analysis. It is related to the cumulative distribution function $F(x)$ which is the probability that a random variable has a value not larger than $x$:

\[
R(x)\,=\,1\,-\,F(x).
\]
The failure rate $\Lambda (x)$ is defined by
\[
\frac{dR}{dx}\ =\,-\Lambda(x)\cdot R(x),
\]
which gives
\begin{equation}
\Lambda (x)=\, -\frac{d\,\ln R}{dx}.\label{4} 
\end{equation}\\


We note that $\Lambda$ is the conditional probability for an event to occur in the interval $(x,x+dx)$ provided it has not occurred until $x$. $\Lambda$ is a very important quantity because it contains the physics. For a constant failure rate, e.g., we obtain the exponential decay of $R$ known from radioactive decay. For a so called Rayleigh distribution $R(x) = \exp (-x^2)$ we have $\Lambda = 2x$.
With these tools we now evaluate first the statistics of the turbulent phases.

\section{Statistics of the Turbulent Phases}
\label{sec:3}

\begin{figure*}
  \includegraphics[width=1\textwidth]{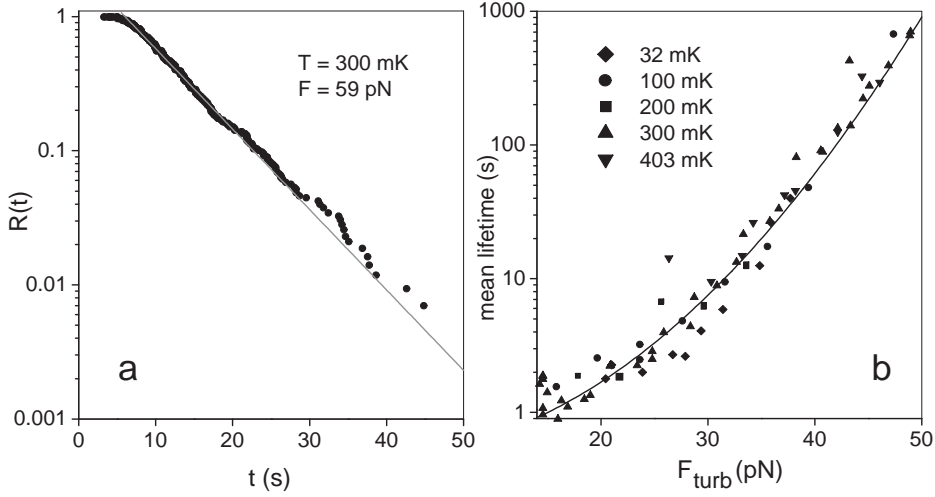}
\caption{(From Ref.3) Statistical analysis of the lifetimes of the turbulent phases. a) The exponential decay of the reliability function $R(t)$ gives the mean lifetime.   b) The mean lifetimes measured at different temperatures and driving forces as a function of the turbulent drag force. The solid line is a fit of an exponential increase with the square of the driving force, see Eq.3.}
\label{fig:3}       
\end{figure*}

In Fig.3a the distribution $R(t)$ of the lifetimes of the turbulent phases is shown for a particular drive and temperature. From the straight line behavior we obtain an exponential decay 
\begin{equation}
R(t)=\exp(-\nu t)\label{6} 
\end{equation}
with $1/\nu = <\!t\!>$ being the mean lifetime. Thus the failure rate $\Lambda= \nu$ is constant, i.e., there is no history dependence in the probability for the decay of a turbulent phase. The values for the mean lifetime are independent of temperature and are found to collapse to a universal drive dependence if the strongly temperature dependent phonon drag $\lambda v_t$ at the velocity $v_t$ is subtracted from the external driving force. ($\lambda$ is the linear drag coefficient due to phonon scattering.) This is reasonable because the phonon scattering is probably not affected by the turbulence. Obviously, the mean lifetime $<\!t\!>$ depends only on the turbulent drag force $F_{turb} = F-\lambda v_t$, see Fig.\,3b. The rapid increase of the lifetime can be described by an exponential growth

\begin{equation}
<\!t\!> = t_0 \,\,\exp\,(F_{turb}/F_1)^2 
\end{equation}
with $t_0 = 0.5$\,s and $F_1=18.3$\,\,pN.\\ 

From Fig.1 we find the turbulent drag force being given by $F_{turb} = \gamma \,(v^2 -v_c^2)$, where $\gamma = c_D\rho_s(T)\pi r^2/2 $ and $c_D = 0.36$ for our sphere \cite{PRL1,OVL}. For simplicity, from here on we shall omit the subscript "turb", i.e., assuming that the linear drag force is always subtracted from the external driving force (but keeping in mind that, strictly speaking, a correction factor of 8/3$\pi \approx$ 0.85 is needed when comparing the driving force with the nonlinear turbulent drag force \cite{BDV}): $F(v) = \gamma (v^2-v_c^2)$. Writing $\gamma v_c^2 = F_0$ gives $F(v)/F_0 = (v^2-v_c^2)/v_c^2$ and from $L = (v/\kappa)^2$, where the amplitude of the vortex density $L$ is the steady state solution of the Vinen equation\footnote{Here we apply the Vinen equation to oscillatory flow provided that the relaxation time of $L$ is short compared to the period of the oscillations, for details, see \cite{JLTP153}.}\cite{kolya}, we have: 
\begin{equation}
\frac{F}{F_0} = \frac{L-L_c}{L_c} \equiv \frac{\Delta L}{L_c},\\ 
\end{equation}
\\
where $L_c = (v_c/\kappa)^2\sim (\sqrt{\kappa \,\omega} /\kappa)^2 \sim \omega/\kappa$. We note that $L_c$ is the lowest vortex density for turbulence to exist and is independent of the drive. Any fluctuation of $L$ that drops below $L_c$ will cause a breakdown of turbulence. For a typical drive $F$ = 30 pN (see Fig. 3b) and a typical $F_0$ = 450 pN (at 120 Hz) we have\\  
\begin{equation}
\frac{\Delta L}{L_c} = \frac{F}{F_0} = 6.7 \cdot 10^{-2}\;\; \mbox{or}\;\; \frac{\Delta L}{L} = \frac{F}{F + F_0} = 6.3 \cdot 10^{-2}.
\end{equation}
This means, that in this case a fluctuation larger than 6.3\% of $L$ will cross the level of $L_c$ downwards and, therefore, cause a breakdown of turbulence. How often does that happen?\\

In the following a model is being presented that describes both the exponential distribution of $R(t)$, see Eq.2, and the dependence of the mean lifetime $<\!t\!>$, see Eq.3. Let us consider a stationary normal process as sketched in Fig.4.
\begin{figure}
  \includegraphics{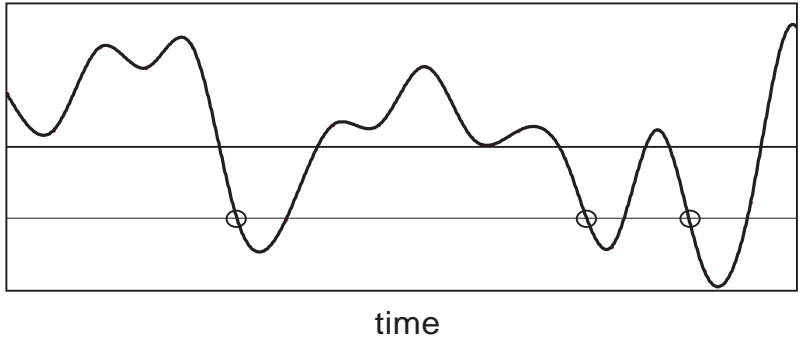}
\caption{(From Ref.3) Sketch of a stationary normal process. The statistical fluctuations around a stationary mean are assumed to have a normal distribution. The down crossings of a given level below the mean are indicated.}
\label{fig:4}       
\end{figure}
The properties of level crossing by a random process are mathematically very well investigated \cite{Rootzen}. First of all, the down crossings constitute a Poisson process on the time axis and, therefore, the time intervals are exponentially distributed. We postulate that the lifetime of a turbulent phase is given by the first down crossing. Consequently, the times of the first down crossings are also exponentially distributed when the system is restarted many times with a new turbulent phase. This explains the exponential distribution of $R(t)$, see Eq.2. It is also plausible, that an increase of the drive will shift the level $L$ further up and, hence, a larger negative fluctuation is needed for a first down crossing below $L_c$ and, consequently, the lifetimes of the turbulent phases will grow. For a quantitative analysis we apply Rice's formula \cite{Rootzen,Rice} for the average number $<N>$ of down crossings per unit time of a level $C$ below the mean of a stationary normal process $\xi(t)$ with zero mean and variance $\sigma^2$, namely,

\begin{equation}
<\!N\!>\,=\,\frac{1}{2\pi}\sqrt{-r''(0)}\exp-\left(\frac{C}{\sqrt{2} \sigma}\right)^2,\label{3}
\end{equation}
where $r''(0)$ is the second derivative of the autocorrelation function of the random process at the origin that is related to the second spectral moment by 
\begin{equation}
-r''(0)\,=\,\int_{0}^\infty \omega^2S(\omega)\,d\omega, \label{4}
\end{equation}
with $S(\omega)$ being the spectral density function of the process. The exponential dependence of $<\!N\!>$ on the level $C$ reflects the assumed normal probability density function of the underlying process $f(\xi)\,=\,(1/\sqrt{2\pi}\sigma)\exp(-\xi^2/2\sigma^2)$. From the measurements of the average time $<\!t\!>$ of the first downcrossing we have the condition
\[
<\!N\!>\cdot<\!t\!>\,=\,1
\]
and therefore
\begin{equation}
<\!t\!>\,=\,\frac{1}{<\!N\!>}\,=\,\frac{2\pi}{\sqrt{-r''(0)}}\exp \left(\frac{C}{\sqrt{2}\sigma}\right)^2 = t_0 \,\,\exp\left(\frac{F}{F_1}\right)^2.\label{5}
\end{equation}
Writing
\[\frac{F}{F_1} = \frac{F}{F_0} \cdot \frac{F_0}{F_1} = \frac{\Delta L}{L_c}\cdot \frac{F_0}{F_1} = \frac{\Delta L}{L_1},\;\; \mbox{where} \;L_1 = L_c\cdot\frac{F_1}{F_0},\] 
we can compare the exponents: $C \equiv \Delta L$ and $\sqrt{2}\,\sigma \equiv L_1$ and finally we obtain for the  width $\sigma$ of the Gaussian fluctuations 

\begin{equation}
\sigma = \frac{L_1}{\sqrt{2}} = \frac{L_c\; F_1}{\sqrt{2}\; F_0}\;\; \mbox{and}\;\; \frac{\sigma}{L} = \frac{L_1}{\sqrt{2}\,L} = \frac{F_1}{\sqrt{2}\,(F+F_0)}.\\
\end{equation}
For a drive $F$ = 30 pN and for $F_0$ = 450 pN and $F_1$ = 18 pN we find $\sigma /L$ = 2.7 $ \cdot10^{-2}$. 
In Fig.5 this numerical example is sketched graphically.\\

The prefactor $t_0$ = 0.5 s in Eq.8 is interpreted as twice the average time between two consecutive crossings of level $L$\, \cite{Rice}, i.e., there are 4 crossings per second. Hence, the average frequency of the fluctuations is 1.5 Hz, which is surprisingly low.\\


\begin{figure*}
 \begin{center}
\includegraphics[width=0.6\textwidth]{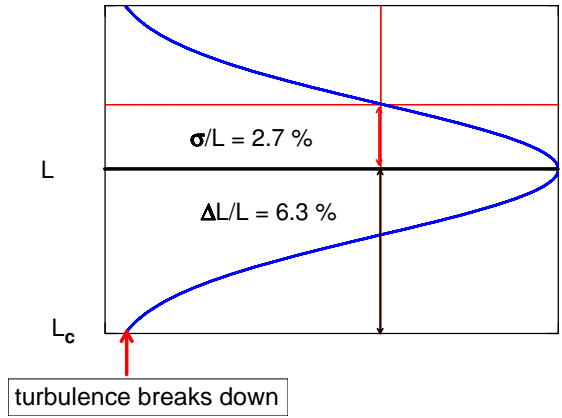}
\caption{(Color online) At a drive of $F$ = 30 pN the vortex density $L$ is 6.3\% above the critical value $L_c$ and the relative width of $L$ is 2.7\%. All fluctuations crossing the level $L_c$ will cause a breakdown of the turbulent state. With increasing drive the level $L$ increases linearly while $L_c$ and the width $\sigma$ remain unchanged. Consequently, the lifetimes of the turbulent state grow rapidly.}
\label{fig:5}       

\end{center}
\end{figure*}

We now have achieved a quantitative interpretation of the statistical properties of the turbulent phases that no longer requires qualitative dimensional considerations as earlier \cite{PRL2}. Our model of a stationary normal process on which the Rice formula is based, allows a detailed analysis of our experimental results. An interesting implication is that although the lifetimes of the turbulent phases grow rapidly with increasing drive they never diverge. That means that turbulence is inherently unstable. When the finite time of experimental observation is shorter than the lifetime a turbulent phase, turbulence only appears to be stable. An analogous situation has recently been observed in classical pipe flow  where turbulence is shown to have lifetimes that grow very rapidly with the Reynolds number without diverging at some critical value \cite{Hof}.\\

\section{Statistics of the Phases of Potential Flow}
\label{sec:4}

\begin{figure*}
  \includegraphics[width=1\textwidth]{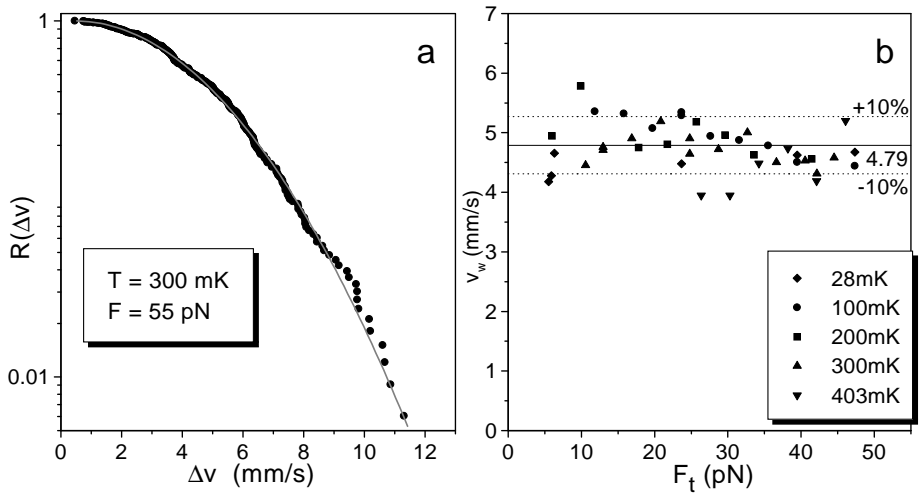}
\caption{Statistical properties of the laminar phases. a) The reliabilty function $R(\Delta v) = \exp(-(\Delta v /v_w)^2)$ is a Rayleigh distribution. b) The fitting parameter $v_w$ is independent of temperature and drive.}
\label{fig:6}       
\end{figure*}

In Fig.6 the statical properties of the laminar phases are displayed. The reliability function $R$ is shown in Fig.6a as a function of the excess velocity amplitude $\Delta v = v - v_t$ above the velocity $v_t$ at turbulent drag at a particular temperature and drive. The data can be fitted by a Rayleigh distribution $R(\Delta v) = \exp(-(\Delta v /v_w)^2)$ and the fitting parameter $v_w$ is shown in Fig.6b to be independent of temperature and driving force (phonon drag being subtracted again). We find, however, that $v_w$ has some frequency dependence. For further discussion we are changing variables now by considering oscillation amplitudes $\Delta a = \Delta v/\omega$ rather than $\Delta v$, i.e., we set $a_w \equiv v_w/\omega$. It is a property of the Rayleigh distribution that the fitting parameters $v_w$ and $a_w$ are the rms values of the reliability functions $R$. Although the range of frequencies of our data is rather limited, a power law fit to the data in Fig.7 seems to indicate an $\omega ^{-3/2}$ dependence of $a_w$. The rms values $a_w$ drop from 9 $\mu$m at 114 Hz to 3.4 $\mu$m at 212 Hz. This means that at high frequencies the laminar phases are breaking down at lower excess amplitudes $\Delta a$ than at low frequencies.

Discussing these results the following questions have to be considered: 1. Why do we observe a Rayleigh distribution of the excess amplitudes $\Delta a$? 2. Why has the rms value $a_w$ this frequency dependence?

\begin{figure*}
\begin{center}
  \includegraphics[width=0.4\textwidth]{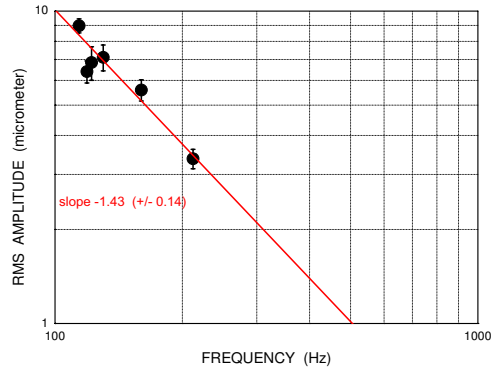}
\caption{(Color online) The rms amplitudes $a_w$ for six different oscillation frequencies. Each data point is obtained from the average of several individual time series recorded at different temperatures and driving forces. The error bars are calculated from the standard deviation. A power law fit that takes into account the error bars indicates a slope of -1.43 $\pm$ 0.14.}
\label{fig:7}       
\end{center}
\end{figure*}

The answer to the first question is made plausible by considering the failure rate $\Lambda$ of the Rayleigh distribution 
$\Lambda = -d$ ln\,$R$\,/\,$d\,\Delta a = 2\,\Delta a /\, a_w^2 $ that is proportional to $\Delta a$. As the sphere increases its oscillation amplitude during a laminar phase it will eventually encounter remanent vortices that were  trapped when the preceeding turbulent phase has broken down. The collision of the sphere with these vortices will give rise to a breakdown of the potential flow. (A similar observation has been made recently by the Osaka group with a vibrating wire colliding with vortex rings causing the flow to become turbulent\cite{Yano}). The probability for this event to occur is proportional to $\Delta a$. This situation is analogous to the hysteresis observed at higher temperatures where during the initial up-sweep of the driving force (after the cell has been filled with helium) the critical velocity $v_c$ can be exceeded considerably before the transition to turbulence finally occurs because the large normal fluid drag at temperatures above ca. 0.5 K limits the excess amplitude and hence the failure rate. A substantial increase of the drive is needed for the laminar phase to break down. Therefore, the switching phenomenon is replaced by the hysteresis \cite{OVL,JLTP158}. It also fits nicely into this picture that metastable potential flow can be observed when the maximum oscillation amplitude due to phonon drag is reached. In this case the spacing of the remanent vortices is larger than the maximum amplitude of the sphere, no vortices can be reached and, hence, potential flow is metastable. Only because of vorticity generated by the absorption of particles from natural radioactive decay the mean lifetime of these metastable states of superfluid potential flow is limited to 25 minutes \cite{OVL,BDV}.

At present, the second question can be discussed only on a qualitative level. The fundamental length scale of oscillatory superfluid flow is determined by $\sqrt{\kappa/\omega}$ (analogous to the viscous penetration depth of oscillating classical flow). This is so for various reasons. Firstly, the oscillation amplitude at the critical velocity $v_c \sim \sqrt{\kappa \omega}$ is $a_c = v_c/\omega \sim \sqrt{\kappa/\omega}$. Secondly, the intervortex spacing at $v_c$ is $l_c = 1/ \sqrt{L_c} \sim \sqrt{\kappa/\omega}$ (and hence scales as $a_c$). Moreover, and very interestingly, also from the dispersion  relation of Kelvin waves, namely
$\omega (k) = \kappa \,k^2 $ln$\left(2 /k a_0\right)/4\pi$,
we obtain a wavelength $\lambda = 2\pi/k \sim \sqrt{\kappa/\omega}$ when considering the logarithmic term to be approximately constant ($\approx$ 12 in our case). It is also interesting to note that the group velocity $v_g \sim \sqrt{\kappa \,\omega}$ scales as $v_c$. Therefore, it appears plausible that Kelvin waves traveling along a vortex that is attached to the oscillating sphere will be excited very efficiently.   
On dimensional grounds we set $a_w \propto \omega ^{-3/2} \propto {(\kappa/\omega)^{3/2}}$ to eliminate the time. It follows that another length scale, independent of $\omega$, must be involved in order to get the dimensions right, probably the radius $r$ of the sphere: $a_w \sim (\kappa/\omega)^{3/2}/r^2.$ The average displaced liquid volume is $\Delta V \approx \pi r^2 \cdot a_w$ so that $\Delta V \sim (\kappa/\omega)^{3/2}$. Unfortunately, it remains unclear whether $\Delta V$ is, by any means, related to $a_c^3$ or $l_c^3$ or $\lambda^3$?

There is a possible alternative way of reasoning based on diffusion of vorticity \cite{Nemi}. But it remains unclear as well, how in detail the intervortex spacing at the end of a turbulent phase increases from $l_c$ to the average  value $l_c + a_w$ by means of a diffusion process with $a_w$ to depend on the period $\tau$ of the oscillation: $a_w \propto \omega ^{-3/2} \propto \tau^{3/2}$.

\section{Conclusion}
\label{sec:5}

We have a achieved detailed picture of the lifetime of the turbulent phases in terms of a fluctuating vortex density $L$. We have found that the fluctuations of $L$ have a normal distribution and have determined the variance. The lifetimes of the turbulent states increase very rapidly with increasing drive but never diverge. Thus, turbulence is  inherently unstable in our experiment. Our picture of the breakdown of potential flow by some remanent vorticity is not yet complete. While the occurrence of the Rayleigh distribution of the excess amplitudes $\Delta a$ is plausible, the dependence of the rms value on the oscillation frequency cannot be explained yet. Also we do not have a sufficient set of data for a definite analysis of this effect. It would be very helpful to have a theory of oscillating superflow which up to date is not yet available.

\begin{acknowledgements}
The data analyzed in this work were obtained together with my former co-workers Hubert Kerscher \cite{Kerscher} and Michael Niemetz \cite{Niemetz}, both of them did a marvelous job when working with me. With Risto H\"anninen (Aalto University, Helsinki) I had a fruitful and pleasant co-operation on the critical velocities of oscillating superflows. I have benefited very much from the hospitality and the stimulating atmosphere during numerous visits to the Helsinki low temperature group lead by Matti Krusius. All of these friends deserve my most sincere gratitude. The workshop "Vortices 2010" where this work was presented, was extremely interesting and pleasant. I am thanking the organizers at Aalto University for inviting me to participate. Eyjafjallaj\"okull made my return trip home an unforgettable odyssey by bus through the Baltic states and Poland.  
\end{acknowledgements}
\bigskip


\end{document}